\documentclass[preprint,prd,aps,nofootinbib,amssymb]{revtex4}
\usepackage{graphicx}
\usepackage{epsfig,amssymb,amsmath}
\newcommand{\beq}{\begin{equation}}
\newcommand{\eeq}{\end{equation}}
\newcommand{\bea}{\begin{eqnarray}}
\newcommand{\eea}{\end{eqnarray}}
\newcommand{\bear}{\begin{array}}
\newcommand {\eear}{\end{array}}
\newcommand{\bef}{\begin{figure}}
\newcommand {\eef}{\end{figure}}
\newcommand{\bec}{\begin{center}}
\newcommand {\eec}{\end{center}}
\newcommand{\non}{\nonumber}

\newcommand{\ds}{\displaystyle}

\def\EQ#1{Eq.~(\ref{#1})}

\def\REF#1{(\ref{#1})}
\def\GEV#1{10^{#1}{\rm\,GeV}}

\def\lrf#1#2{ \left(\frac{#1}{#2}\right)}
\def\lrfp#1#2#3{ \left(\frac{#1}{#2} \right)^{#3}}

\newcommand{\del}{\partial}

\begin{document}
\draft
\tighten
\preprint{
TU-979, IPMU14-0269
}
\title{\large \bf
Hidden axion dark matter decaying through
 mixing with QCD axion and the $3.5$ keV X-ray line
}
\author{
    Tetsutaro Higaki\,$^a$\footnote{email: thigaki@post.kek.jp},
    Naoya Kitajima\,$^b$\footnote{email: kitajima@tuhep.phys.tohoku.ac.jp},
    Fuminobu Takahashi\,$^{b,c}$\footnote{email: fumi@tuhep.phys.tohoku.ac.jp}
    }
\affiliation{
 $^a$ Theory Center, KEK, 1-1 Oho, Tsukuba, Ibaraki 305-0801, Japan \\
 $^b$ Department of Physics, Tohoku University, Sendai 980-8578, Japan\\
 $^c$ Kavli Institute for the Physics and Mathematics of the Universe (WPI), 
 TODIAS, University of Tokyo, Kashiwa 277-8583, Japan    }

\vspace{2cm}

\begin{abstract}
Hidden axions may be coupled to the standard model particles through a kinetic or mass mixing with
QCD axion.  We study a scenario in which a hidden axion constitutes a part of or the whole
of dark matter and decays into photons through the mixing,  explaining the $3.5$\,keV X-ray line signal.
Interestingly, the required long lifetime of the hidden axion dark matter can be realized
 for the  QCD axion decay constant at an intermediate scale, if the mixing is sufficiently small. 
In such a two component dark matter scenario, the primordial density perturbations of the
hidden axion can be highly non-Gaussian, leading to a possible dispersion in the X-ray line
strength from various galaxy clusters and near-by galaxies. We also discuss how the parallel and
orthogonal alignment of two axions affects their couplings to gauge fields. 
\end{abstract}

\maketitle

\section{Introduction}
Recently two groups independently reported a detection of an unidentified X-ray line at energy about $3.5$~keV 
in the stacked XMM-Newton spectrum of 73 galaxy clusters~\cite{Bulbul:2014sua} and in the out-skirts of
the Perseus cluster as well as in the center of the Andromeda galaxy (M31) ~\cite{Boyarsky:2014jta}. 
The detection of the X-ray line signal has led to much excitement as it could be due to decaying dark matter (DM)
such as sterile neutrinos~\cite{Dolgov:2000ew,Ishida:2014dlp,Abazajian:2014gza}.
Furthermore, an X-ray line at energy $3.53$ keV was detected in the XMM-Newton
data of the Galactic center, which is consistent with the DM interpretation of the above 
observations~\cite{Boyarsky:2014ska}.\footnote{Such X-ray line signal was however questioned 
in Ref.~\cite{Riemer-Sorensen:2014yda} where no clear evidence for the line 
was found in the Chandra X-ray data of the Galactic center. 
Also, allowing a high $3.52$\,keV K XVII line flux, no conclusive excess line emission was found in the 
XMM-Newton data of the 
Galactic center, M31 and galaxy clusters  in Ref.~\cite{Jeltema:2014qfa}.
The recent analysis on the stacked dwarf  spheroidal galaxies show no excess which places a tight
constraint on the DM interpretation of the excess~\cite{Malyshev}.
}

After the detection of the $3.5$\,keV X-ray line, there have been proposed various DM models. 
One of the interesting candidates is a pseudo Nambu-Goldstone 
boson or axion~\cite{Higaki:2014zua,Jaeckel:2014qea,Lee:2014xua,Cicoli:2014bfa,Nakayama:2014cza,
Conlon:2014xsa,Conlon:2014wna}.\footnote{
Another similar candidate is a light moduli field decaying into 
photons studied long ago by Kawasaki and Yanagida in a context of gauge mediation~\cite{Kawasaki:1997ah}.
(See also Refs.~\cite{Hashiba:1997rp,Kusenko:2012ch,Nakayama:2014ova}). }
If the axion mainly decays into photons,   the following mass and lifetime are required 
for explaining the $3.5$\,keV X-ray line signal~\cite{Boyarsky:2014jta}
\begin{align}
	\label{mass}
	m_{\rm DM} &\simeq 7.1\,{\rm keV} \\
	\tau_{\rm DM}/r_d& \simeq 4\times 10^{27} - 4\times 10^{28}\,{\rm sec.}
	\label{tau}
\end{align}
where $r_d$ is the fraction of the axion DM in the total DM density,
and this expression is valid for the lifetime  longer than the present age of the Universe,
i.e., $r_d \gtrsim 10^{-10}$.
The light mass can be understood if the shift symmetry of the axion is broken weakly.
Interestingly,  for $r_d \simeq 1$, the required long lifetime implies the decay constant  
of order $10^{14} - 10^{15}$\,GeV,
close to what is expected for the string axions~\cite{Higaki:2014zua}.  
Note that the axion cannot be the QCD axion which solves the strong CP problem,
because the axion mass has to be much lighter for the decay constant given above. 

The required long lifetime does not necessarily point to  physics at high energies close to the GUT scale,
if there is an additional suppression factor in the axion coupling with photons. This is the case 
if the corresponding U(1) symmetry is anomaly-free and the axion is coupled to photons only through 
mass threshold corrections~\cite{Nakayama:2014cza}. Alternatively, the hidden axion may be originally
coupled to only hidden photons, which has a small kinetic mixing with photons~\cite{Jaeckel:2014qea}.

 In this letter, as another way to suppress
the coupling, we consider a hidden axion DM decaying into photons through a small kinetic or mass mixing with 
the QCD axion. Interestingly, our model requires both the QCD and hidden axions, thereby leading to a 
two-component DM scenario. As we shall see later, there are  various interesting implications. 
In particular, the long lifetime (\ref{tau}) can be realized even with the QCD axion decay constant at an intermediate
scale,  independently of the hidden axion decay constant. Therefore it is possible that both the QCD and hidden
axions appear at an intermediate scale. In  field-theoretic axion models, both U(1) symmetries can be
restored during inflation, avoiding the isocurvature constraints. Also, the small kinetic mixing with the
QCD axion couples hidden axions to the visible sector, even if there are no directly couplings.
In general, such couplings can affect the cosmological impact of axions and moduli fields.
We will also discuss how the parallel and orthogonal alignment of the two axions affects the relation between
the symmetry breaking scale and the decay constant. For instance, the QCD axion decay constant can be
much larger than the actual Peccei-Quinn breaking scale. This is an application of the so called 
Kim-Nilles-Peloso (KNP) mechanism~\cite{Kim:2004rp} which was originally 
proposed to realize an effective super-Planckian 
decay constant for successful natural inflation.

The rest of this letter is organized as follows. In Sec.~\ref{sec2} we first provide a set-up for the hidden and QCD
axions with a kinetic or mass mixing, and then show that the hidden axion DM decaying through the mixing with the QCD
axion can explain the $3.5$\,keV X-ray line.   We discuss the abundances of the two axions 
and cosmological implications in Sec.~\ref{sec3}.
The last section is devoted for discussion and conclusions.

\section{Mixing between hidden axion and QCD axion}
\label{sec2}
\subsection{Kinetic mixing}
Let us consider a situation where the QCD axion and the hidden axion, denoted respectively as $a$ 
and $a_H$, coexist.  The axions $a$ and $a_H$ arise as a (pseudo) Nambu-Goldstone
boson  associated with the spontaneous breakdown of Peccei-Quinn U(1)$_{PQ}$ and  hidden U(1)$_H$ 
symmetries, respectively. We assume here that the hidden axion does not have any direct couplings with the 
standard model (SM) particles, and  we introduce a small kinetic mixing between $a$ and $a_H$,  
which, as we shall see,  couples the hidden axion to the SM sector. 

To be concrete, let us introduce two complex scalar fields $\Phi(1,0)$ and $\phi(0,1)$, where the charges 
under U(1)$_H$ and U(1)$_{PQ}$ are shown in the parentheses. We assume that they develop non-zero
vacuum expectation values (VEVs),
\bea
&&\Phi = \frac{F_H + \sigma_H}{\sqrt{2}} e^{i a_H/F_H},~~~
\phi = \frac{F_a + \sigma}{\sqrt{2}} e^{i a/F_a},
\label{LR}
\eea
where $F_H$ and $F_a$ represent the VEV, and $\sigma_H$ and $\sigma$ are the hidden and QCD saxions,
respectively. We assume that the saxions are stabilized with a mass much heavier than the axions, and therefore,
they can be safely integrated out in the low energy. It is possible to consider multiple scalar fields or non-linearly 
realized symmetries,  but the following arguments can be straightforwardly applied to these cases.  

The kinetic mixing between two axions can be induced from the following  operator (or similar ones)\footnote{
In supersymmetry, such kinetic mixing is induced from $K = |\phi|^2 |\Phi|^2/M^2$. The effect of kinetic
and mass mixings in supergravity was studied in detail in Ref.~\cite{Endo:2006tf}.
See also Ref.\cite{Higaki:2011bz}, in which a mixing between the QCD axion 
between the $R$-axion was discussed.
}
\beq
\mathcal{L} _{\rm eff} \supset \frac{1}{ M^2} (\phi^\dagger \del_\mu \phi\, \Phi \del^\mu \Phi^\dagger + {\rm h.c.}),
\eeq
where $M$ is an effective cut-off scale. Such kinetic mixing can be induced by integrating out a heavy field 
charged under both U(1)$_H$ and U(1)$_{PQ}$ symmetries~\cite{Babu:1994id}. Substituting \EQ{LR},
we obtain the kinetic mixing between $a_H$ and $a$ as
\bea
\mathcal{L} _{\rm eff} \supset \epsilon\, \partial_\mu a \partial^\mu a_H
\eea
with
\beq
\epsilon =  \frac{F_a F_H}{2M^2}.
\eeq
Alternatively, such kinetic mixing may be built-in or induced from some strong dynamics 
in the case of  non-linearly realized symmetries.
In the following analysis we take $\epsilon$ as a free constant parameter smaller than unity, but 
it should be kept in mind that $\epsilon$ may not be a constant at high energies;
for instance, in the above example, the kinetic mixing will be reduced at energy scales 
above the mass of the heavy field, and also, the U(1) symmetries may be restored at
high energies.

The quadratic part of the Lagrangian for the axions is given by
\beq
	\mathcal{L} = \frac{1}{2}(\del_\mu a)^2 + \frac{1}{2} (\del_\mu a_H)^2 
	+ \epsilon\, \del_\mu a \del^\mu a_H -\frac{1}{2}m_a^2 a^2 -\frac{1}{2} m_H^2 a_H^2,
	 \label{Lag}
\eeq
where $m_a$ and $m_H$ are the mass of $a$ and $a_H$ respectively,
and $\epsilon$ is a small numerical factor of the kinetic mixing.
Note that the kinetic mixing is allowed by the  shift symmetries of $a$ and $a_H$.
The QCD axion mass,  $m_a$, is 
induced by QCD non-perturbative effects at temperatures below the QCD phase transition,
and given by
\beq
m_a \approx  6.0 \times 10^{-6}  {\rm \, eV} \lrf{\GEV{12}}{f_a},
\eeq
where $f_a$ is the QCD axion decay constant, which is usually comparable to $F_a$,
but could be significantly different as we shall see later.  Here we neglect
the effect of the kinetic mixing on the axion mass (see below). 
Similarly,  the hidden axion mass is induced by non-perturbative effects of hidden gauge interactions,
and it is roughly given by $m_H \sim \Lambda_H^2/f_H$, where $\Lambda_H$ denotes the dynamical
scale  and $f_H$ is the hidden axion decay constant. Here we assume $m_H \gg m_a$
and there is no mixing in the mass term in this basis.\footnote{Even in the presence of
the kinetic and mass mixings, the strong CP problem can be solved by the PQ mechanism.} 
The effect of the mass mixing will be considered next. 
For the moment, we also assume that there is no additional light degrees of freedom 
such as hidden photons. We shall return to this issue later.

To make the kinetic terms canonically normalized, 
we perform the following linear transformation 
\bea
\label{tr1}
a &=& a' - \frac{\epsilon}{\sqrt{1-\epsilon^2}} a_H',\\
a_H &=& \frac{a_H'}{\sqrt{1-\epsilon^2}}
\label{tr2}
\eea
and then,  the Lagrangian is rewritten as
\beq
	\mathcal{L} \;=\; 
	 \frac{1}{2} (\del_\mu a')^2 + \frac{1}{2} (\del_\mu a_H')^2 -\frac{1}{2} m_a^2 a'^2 
	  -\frac{1}{2}\frac{m_H^2 + \epsilon^2 m_a^2 }{1-\epsilon^2} a_H'^2 
	  + \frac{\epsilon}{\sqrt{1-\epsilon^2}} m_a^2 a' a_H',
	\label{Lag2}
\eeq
where the last term represents a mass mixing between $a'$ and $a_H'$.  The induced mass mixing is, however, 
of order $\epsilon \,m_a^2/m_H^2$, which is much smaller than the kinetic mixing as long as 
$m_H^2 \gg m_a^2$. The two mass eigenvalues are given by 
\beq
m_H^2 + \frac{m_H^4}{m_H^2 - m_a^2} \epsilon^2,~~~m_a^2 - \frac{m_a^4}{m_H^2-m_a^2} \epsilon^2
\eeq
up to  corrections of $O(\epsilon^4)$. Thus, the change of the mass eigenvalues is negligibly small 
for $|\epsilon| \ll 1$ and $m_H^2 \gg m_a^2$, and so,
we will neglect the induced mass mixing as well as the change of the mass eigenvalues.
In effect, the kinetic mixing with $|\epsilon| \ll 1$ results in a shift of the axion $a$ by $- \epsilon a_H'$ as in \EQ{tr1}.

\subsection{Mass mixing}
Next we consider a mass mixing between hidden axion and QCD axion. Here we assume that there is
no kinetic mixing.  The mass mixing arises if both $a$ and $a_H$ are coupled to gluons and
 hidden gauge fields. In the low energy, those axions acquire a potential of the cosine form from instanton effects:
\beq
V = \Lambda_H^4 \left(1-\cos\left(n_1 \frac{a_H}{F_H} + n_2 \frac{a}{F_a} \right) \right)
+ \Lambda_{QCD}^4 \left(1-\cos\left(m_1 \frac{a_H}{F_H} + m_2 \frac{a}{F_a} \right)\right),
\eeq
where $n_1$, $n_2$, $m_1$, and $m_2$ are integers which depend on the number of (hidden) quarks
charged under the U(1)$_H$ and U(1)$_{PQ}$ symmetries, and we have dropped the CP phases.
One can clearly see that the two axions are mixed by the mass term in this case. 
To simplify our analysis, we assume that the hidden axion is much heavier than the QCD axion.
Then the two mass eigenstates $\tilde{a}_H$ and $\tilde a$ can be approximately given by
\bea
\tilde a_H &\approx& \ds{\frac{1}{\sqrt{\lrfp{n_1}{F_H}{2}+\lrfp{n_2}{F_a}{2}}}\left(n_1 \frac{a_H}{F_H} + n_2 \frac{a}{F_a}\right)},\\
\tilde a &\approx& \ds{\frac{1}{\sqrt{\lrfp{n_1}{F_H}{2}+\lrfp{n_2}{F_a}{2}}}\left(-n_2 \frac{a_H}{F_H} + n_1 \frac{a}{F_a}\right)},
\eea
up to a correction suppressed by the mass squared ratio. 
Assuming that the same combination of $a$ and $a_H$ couples to both gluons and photons, 
the coupling of the hidden axion $\tilde a_H$ to photons is given by
\bea
{\cal L} &\supset& \frac{\alpha_{\rm EM}}{4\pi} c_\gamma
\left(m_1 \frac{a_H}{F_H} + m_2 \frac{a}{F_a} \right) F_{\mu\nu}\tilde{F}^{\mu\nu},\non \\
&=&  \frac{\alpha_{\rm EM}}{4\pi}c_\gamma \frac{ \left(m_1 n_1 + m_2 n_2 \right) \tilde a_H + 
\left(-m_1 n_2 + m_2 n_1 \right) \tilde a }{\sqrt{(n_1 F_a)^2+(n_2 F_H)^2}}
F_{\mu\nu}\tilde{F}^{\mu\nu},
\eea
where  $\alpha_{\rm EM}$ is the electromagnetic fine structure constant, 
 $c_\gamma$ is a numerical factor which depends on the details of the model, and $F_{\mu \nu}$
is the electromagnetic field strength. Therefore, as expected, the hidden axion mass eigenstate 
is coupled to photons.

Interestingly, the coupling can be suppressed for a certain choice of $(n_1, n_2)$ and $(m_1, m_2)$.
This is a simple application of the KNP mechanism~\cite{Kim:2004rp} to the hidden and QCD axions. 
Here, it is orthogonal alignment (i.e., $m_1 n_1  + m_2 n_2 \approx 0$) that suppresses the coupling. 
For instance, if $m_1 =0$, $m_2 \sim n_2 = O(1)$,  $n_1 \gg 1$ and $F_a \sim F_H$, the effective
decay constant can be enhanced by about $n_1$~\cite{Tye:2014tja,Ben-Dayan:2014zsa}. 
As long as only two axions is considered, one needs a hierarchy among the  integer-valued coefficients,
which requires many heavy (hidden) quarks. Note that $m_1$ and $m_2$ cannot be arbitrarily large
because the number of extra quarks is  constrained by the perturbativity limit of the gauge interactions 
up to the GUT scale. The required hierarchy can be
 relaxed in the presence of multiple axions~\cite{Choi:2014rja,Higaki:2014pja}.
Thus it is possible to suppress the coupling of the hidden axion to photons by mass mixing with
a certain choice of the coefficients, but it requires some hierarchy in parameters or non-trivial extensions.
Therefore we will mainly focus on the kinetic mixing in the next subsection. 

Before closing this subsection, we  note that the couplings of the QCD axion to the SM gauge fields 
can be similarly suppressed for a certain choice of of $(n_1, n_2)$ and $(m_1, m_2)$ (parallel alignment, i.e., 
 $-m_1 n_2  + m_2 n_1 \approx 0$).
To put it another way, the QCD axion decay constant can be
much larger than the actual PQ symmetry breaking scale. This implies that, even if the effective QCD
axion decay constant is close to the GUT scale, the U(1)$_{PQ}$ symmetry can be restored 
at a temperature or inflation scale much lower than the GUT scale.

\subsection{The $3.5$\,keV X-ray line}
The QCD axion has various couplings to the SM particles. Among them, the axion coupling to photons is
particularly interesting from both experimental and cosmological point of view. Hidden axions with couplings
to photons are often called axion-like particles (ALPs)~\cite{Masso:1995tw,Masso:2004cv,Jaeckel:2006xm,Cadamuro:2011fd,Arias:2012az}, and the mixing between the QCD axion and an ALP was studied in Ref.~\cite{Cicoli:2012sz}.  Using the transformation \EQ{tr1}, we obtain
\beq
{\cal L} \supset \frac{\alpha_{\rm EM}}{4\pi} c_\gamma  \frac{a}{f_a} F_{\mu\nu}\tilde{F}^{\mu\nu}
= \frac{\alpha_{\rm EM}}{4\pi} \frac{c_\gamma}{f_a} \bigg( a'-\frac{\epsilon}{\sqrt{1-\epsilon^2}} a_H' \bigg) 
F_{\mu\nu}\tilde{F}^{\mu \nu}.
\label{ALP}
\eeq
Therefore, the kinetic mixing with the QCD axion turns the hidden axion into
an ALP.  Note that the hidden axion is similarly coupled to gluons, weak gauge 
bosons through the kinetic mixing. In the case of the DFSZ axion model~\cite{Zhitnitsky:1980tq,Dine:1981rt},
it is also coupled to the SM fermions.
In the following we drop the prime on $a'$ and $a_H'$ where there is no confusion.

The last term in \EQ{ALP} implies that the hidden axion can decay into two photons with an effective decay constant 
$\sim f_a/\epsilon$, even if there is no direct coupling in the original basis. 
Then, the decay rate of the hidden axion into two photons is calculated as
\beq
	\Gamma_{a_H \to \gamma\gamma} = \frac{\alpha_{\rm EM}^2}{64 \pi^3} \frac{m_H^3}{f_{\rm eff}^2},
\eeq
where $f_{\rm eff}$ is the effective decay constant defined by 
\beq
f_{\rm eff} \equiv \frac{\sqrt{1-\epsilon^2}}{c_\gamma \epsilon} f_a.
\label{feff}
\eeq
Assuming that the hidden axion mainly decays into photons, its lifetime is estimated as
\beq
	\tau_H \simeq 2 \times 10^{28}~{\rm s}~\bigg( \frac{\alpha_{\rm EM}}{1/137} \bigg)^{-2} \bigg( \frac{m_H}{7~{\rm keV}} \bigg)^{-3}
	\bigg( \frac{f_{\rm eff}}{5 \times 10^{14}~{\rm GeV}} \bigg)^2.
\eeq
Thus, if the hidden axion makes up a fraction $r_d$ of the total DM, the $3.5$\,keV X-ray line signal can 
be explained by $f_{\rm eff} \simeq \sqrt{r_d}\, \GEV{14-15}$ and $m_H \simeq 7$\,keV.
In particular, if the hidden axion constitutes all the DM, the required value of $f_{\rm eff}$ is of order $\GEV{14-15}$.
Interestingly such large decay constant can be induced by a combination of the small kinetic mixing and
the QCD axion decay constant at an intermediate scale, e.g. $\epsilon = {\cal O}(10^{-5})$ and 
$f_a \sim 10^{10} ~ {\rm GeV}$. If the fraction of the hidden axion DM is smaller,   $\epsilon$ needs to be larger
with a fixed $f_a$. 
Since $r_d$ is bounded below, $r_d \gtrsim 10^{-10}$, the effective decay constant $f_{\rm eff}$ can be as small as
about $\GEV{10}$, for which a large kinetic mixing $\epsilon \sim 1$ is required. 
Thus,  the observed X-ray line signature at energy
$3.5$\,keV can be explained by physics at intermediate scales. As we shall discuss in the next
section, this is important for reducing the axion abundance and avoiding the isocurvature constraints.

\section{Hidden and QCD axion DM}
\label{sec3}
In this section, first we estimate the abundances of the hidden and QCD axions, both of which contribute to DM.
As we shall see, the production mechanism for the two axions is quite similar. Then we discuss their cosmological
implications.

\subsection{DM abundance}
\subsubsection{Hidden axion}
First,  suppose that the U(1)$_H$ is spontaneously broken during inflation. Then
the $a_H$ starts to oscillate when the Hubble parameter becomes comparable to $m_H$. 
The abundance of the coherent oscillations is given by
\bea
\Omega_H^{\rm (osc)} h^2 \simeq 
\left\{
\bear{cc}
\ds{0.1 \lrfp{g_*(T_{\rm osc})}{106.75}{-\frac{1}{4}} \lrfp{m_H}{7\,{\rm keV}}{\frac{1}{2}} \lrfp{f_H}{\GEV{11}}{2} \theta_H^2}
	& ~~~{\rm for}~~~T_R > T_{\rm osc} \\
	&\\
	\ds{0.05 \lrf{T_R}{\GEV{5}} \lrfp{f_H}{3 \times \GEV{11}}{2} \theta_H^2}
	& ~~~{\rm for}~~~T_R < T_{\rm osc} 
\eear
\right.,
\label{OHo}
\eea
where $h$ is the reduced Hubble constant, $\theta_H = a_{H}^{\rm (ini)}/f_H$  is the initial misalignment angle, and
$T_R$ is  the reheating temperature. Here we have defined $T_{\rm osc}$ as
\beq
	T_{\rm osc} \simeq 2 \times 10^{6}~{\rm GeV} \bigg( \frac{g_*(T_{\rm osc})}{106.75} \bigg)^{-1/4} \bigg( \frac{m_H}{7~{\rm keV}} \bigg)^{1/2},
\eeq
where $T_{\rm osc}$ is the temperature at which the hidden axion starts to oscillate
in the radiation dominated era, and  $g_*$ counts the relativistic degrees of freedom at $T=T_{\rm osc}$.
Note that no fine-tuning of the initial misalignment angle is necessary for the hidden axion to become
the dominant DM component, if $f_H$ is at intermediate scale. As we have seen in the previous section,
such value of $f_H$ is consistent with the $3.5$\,keV X-ray line, 
because the lifetime is determined by the small kinetic mixing and the
QCD axion decay constant, independently of $f_H$. Similar observation concerning the fine-tuning problem was 
made in Ref.~\cite{Jaeckel:2014qea}.

If either the reheating temperature or the Hubble parameter during inflation is higher
than the decay constant, the U(1)$_H$ symmetry is likely restored during or after inflation.\footnote{
This is not the case if $\Phi$ acquires a negative Hubble-induced mass term during inflation.
}
When the U(1)$_H$ becomes spontaneously broken some time after inflation, 
cosmic strings are formed. When the Hubble parameter becomes comparable to $m_H$,
the axion coherent oscillations as well as domain walls are formed. Therefore, there are
three contributions to the present hidden axion abundance; coherent oscillations and the collapse of
strings and domain walls. Note that the axions produced from these sources becomes non-relativistic
soon after production, and they behave as cold dark matter. 
The hidden axion abundance from coherent oscillation is obtained by replacing 
the initial misalignment angle $\theta_H^2$ with $\pi^2 c_{H, {\rm ann}} /3$, where
$c_{H, {\rm ann}}$ is a factor of order unity which represents an anharmonic 
effect~\cite{Turner:1985si,Lyth:1991ub}. Contributions from the collapse of strings and domain walls were
estimated by numerical simulations~\cite{Hiramatsu:2012gg}. The total $a_H$ abundance from
these contributions  is  given by
\bea
\Omega_H h^2 \simeq 
(0.11 \pm 0.04) \lrfp{g_*(T_{\rm osc})}{106.75}{-\frac{1}{4}} \lrfp{m_H}{7\,{\rm keV}}{\frac{1}{2}} \lrfp{f_H}{\GEV{10}}{2}
~~~{\rm for}~~~T_R > T_{\rm osc},
\label{OHr}
\eea
where  we assume that there is a unique vacuum since otherwise 
domain walls would be stable and remain until present,   a cosmological catastrophe. 
Here we have assumed $T_R > T_{\rm osc}$, since we have simply scaled the results 
of Ref.~\cite{Hiramatsu:2012gg}, where radiation dominated Universe was assumed. 
The $a_H$ abundance in the case of $T_R  < T_{\rm osc}$ can be roughly estimated by multiplying 
the above result with $T_R/T_{\rm osc}$.

In addition, the hidden axions can be produced thermally from its couplings to the SM gauge bosons 
induced by the kinetic mixing.  The thermally produced axions contribute to warm DM,
and their abundance is obtained based on the result for the QCD axion~\cite{Turner:1986tb,Masso:2002np,Graf:2010tv,Salvio:2013iaa};
\beq
	\Omega_H^{\rm (th)} h^2 \simeq 3 \times 10^{-6} \bigg( \frac{\gamma}{0.01} \bigg) \bigg( \frac{106.75}{g_*} \bigg) \bigg( \frac{m_H}{7~{\rm keV}} \bigg)
	\bigg( \frac{5 \times 10^{14}~{\rm GeV}}{f_{\rm eff}} \bigg)^2 \bigg( \frac{T_R}{10^{8}~{\rm GeV}} \bigg),
\eeq
where $\gamma$ is some numerical factor smaller than unity, and $f_{\rm eff} \simeq f_H/\epsilon$ (cf. \EQ{feff}).
Note that the couplings of the hidden axion to the SM gauge fields are suppressed by the kinetic 
mixing $\epsilon$.
Thus, the thermally produced hidden axions account for only a fraction of the all DM.
If the hidden gauge sector is reheated by the inflaton decay, there may be
additional contributions, which however we do not consider here.

\subsubsection{QCD axion}
The production processes of the QCD axion have been studied extensively in the literature.
Here we summarize the results. The coherent oscillations of the QCD axion are 
produced by the initial misalignment mechanism, and its abundance is given by~\cite{Turner:1985si,Bae:2008ue}
\beq
\Omega_a^{(\rm osc)} h^2 \simeq 0.18\, \theta_a^2 \lrfp{f_a}{\GEV{12}}{1.19} \lrf{\Lambda_{\rm QCD}}{400 \,{\rm MeV}},
\eeq
where $\theta_a = a_i/f_a$ is the initial misalignment angle for the QCD axion. If the U(1)$_{PQ}$ symmetry
is restored during or after inflation, on the other hand, there are additional contributions from cosmic strings and domain
walls which annihilate when the axion starts to oscillate. The total abundance is given by~\cite{Hiramatsu:2012gg}
\beq
	\Omega_a h^2 \simeq (8.4 \pm 3.0) \times \bigg( \frac{f_a}{10^{12}~{\rm GeV}} \bigg)^{1.19} \bigg( \frac{\Lambda_{\rm QCD}}{400~{\rm MeV}} \bigg).
\eeq
Note that those axions become non-relativistic soon after production, and so, they contribute to cold dark matter.

The QCD axions are also thermally produced and they contribute to hot dark matter. 
Their abundance is given in Refs.~\cite{Turner:1986tb,Masso:2002np,Graf:2010tv,Salvio:2013iaa}.
For the QCD axion decay constant satisfying the SN1987A 
bound, $f_a \gtrsim \GEV{9}$~\cite{Mayle:1987as,Raffelt:1987yt,Turner:1987by}, however, the abundance
is negligibly small, $\Omega_a^{\rm (th)} h^2 \lesssim 10^{-5}$.

\subsection{Cosmological implications}
As we have seen above, both the hidden axion and the QCD axion can make up a significant
fraction of the total DM. Here we discuss various cosmological implications of our scenario.

First, it is possible to explain the $3.5$\,keV X-ray line signal by the decay constants at an
intermediate scale. Thus, the hidden axion need not to be identified with string axions.
One of the advantages of considering field-theoretic axion models is that the corresponding
U(1) symmetries can be restored during or after inflation. Then, since no axions are present
during inflation, there will be no axion isocurvature perturbations, thus avoiding the tight
isocurvature constraint~\cite{Higaki:2014zua}. The isocurvature constraint becomes extremely 
severe if the inflation scale is close to the GUT scale as suggested by the BICEP2 result~\cite{Ade:2014xna} (
see also Refs.~\cite{Higaki:2014ooa,Marsh:2014qoa,Visinelli:2014twa}).

Secondly, our scenario necessarily  leads to a two-component DM scenario. In an extreme case,
the fraction of the hidden axion DM, $r_d$, can be much smaller than unity. For instance, 
this is the case if $f_H$ is smaller than the intermediate scale (cf. Eqs.~\REF{OHo} and \REF{OHr}).
Then, it is possible that the hidden axion is produced mainly by quantum fluctuations during 
inflation~\cite{Suyama:2008nt}.
Specifically, the abundance is approximately given by \EQ{OHo} with $\theta_H$  replaced with $H_{\rm inf}/2\pi F_H$,
where $H_{\rm inf}$ is the Hubble parameter during inflation. In this case, the  hidden axion abundance becomes independent of $F_H$ and its quantum fluctuations generate non-Gaussian CDM isocurvature 
perturbations~\cite{Kawasaki:2008sn,Langlois:2008vk,Kawakami:2009iu,
Langlois:2011zz,Langlois:2010fe,Kobayashi:2013nva}. To be consistent with CMB and large-scale structure
observations, the fraction of the hidden axion DM is bounded above
as $r_d \lesssim 10^{-5}$~\cite{Hikage:2008sk,Hikage:2012be}. Therefore there is a range of
$10^{-10} \lesssim r_d \lesssim 10^{-5}$, where the $3.5$\,keV X-ray line can be explained by
the hidden axion whose primordial density perutrbation is highly non-Gaussian. Interestingly,
such non-Gaussian hidden axion leads to a variation of the fraction of the hidden axion DM in each astrophysical
object. (Note that $r_d$ mentioned above is the averaged fraction of the hidden axion DM.)
Therefore, the observed X-ray line strength from various galaxy clusters and near-by galaxies is 
expected to vary by an $O(1)$ factor. It is then possible to relax or avoid a constraint on the decaying DM
interpretation based on non-detection of the excess from a single source or several sources close to one another,
by assuming that the hidden axion DM density happens to be smaller than the average in the neighborhood of those
objects. Note however that it is difficult to relax the constraint by more than one order of magnitude and to avoid the
constraints based on stacked observations because it would require too much fine-tuning.

\section{Discussion and conclusions}

The existence of Nambu-Goldstone bosons or axions may be ubiquitous in nature.
If so, axions are a plausible candidate for DM since the longevity can be naturally
understood from their light mass. Then, one of them may decay into photons explaining 
the $3.5$\,keV X-ray line signal.
In this letter we have proposed a scenario in which the coupling of the hidden axion DM to photons 
is induced from its mixing with the QCD axion. We have shown that the coupling can be 
suppressed by the small kinetic mixing or alignment of two axions, and the detected
$3.5$\,keV signal can be explained even with  the U(1)$_H$ and U(1)$_{PQ}$ symmetry breaking scales
at intermediate scale. The advantages of considering field-theoretic axions at intermediate scale
are (i) the isocurvature constraints can be avoided if the U(1)$_H$ and U(1)$_{PQ}$
 symmetries are restored; (ii) no fine-tuning of the misalignment angle is needed to generate
 a right amount of the hidden axion DM.  In addition, in our two-component DM scenario,
 the hidden axion DM is allowed to have large non-Gaussianity as long as its DM fraction is
 in the range of $10^{-10} \lesssim r_d \lesssim 10^{-5}$. Then the hidden axion DM fraction $r_d$
 is expected to vary by a factor of $O(1)$ in astrophysical objects. 
 This may lead to an observable dispersion of the X-ray line strength in various galaxy clusters and near-by galaxies.
 
 We have also pointed out that the KNP mechanism can be applied to our scenario. For instance,
 the hidden axion coupling to photons can be suppressed by (approximate) orthogonal alignment of the two axions.
 Similarly, the QCD axion decay constant can be much larger than the actual PQ symmetry breaking scale
 if the two axions are approximately aligned parallel to each other. By making use of this effect, for instance,
 it is possible that the U(1)$_{PQ}$ symmetry is restored at temperatures or the inflation scale much lower
 than the axion decay constant.
 
 The hidden axion may be naturally embedded in the parallel world or the mirror world. Then it may be possible
that both the kinetic mixing that between photons and dark photons~\cite{Jaeckel:2014qea} and that
between QCD axions and hidden axions are present.  In this case, the hidden axion may decay mainly into hidden photons.

So far we have considered a linear realization of the U(1)$_H$ symmetry. Here
we mention a case in which the hidden axion originates from an antisymmetric tensor field
in the string theory. Through a compactification of extra dimension, 
the decay constant $f_H$ of the hidden axion will be of order the string scale 
around the GUT scale.\footnote{
A lower string scale can be realized in e.g. the LVS scenario~\cite{Balasubramanian:2005zx} where
there is a light string moduli field, which leads to a cosmological moduli problem.
}
Importantly, there naturally appears a kinetic mixing between 
the QCD axion and the hidden one through gravitational interactions.
To be specific,  we consider a K\"ahler potential of the form,
\beq
K = \frac{f_H^2}{2} (A+A^\dag)^2 + Z(A+A^{\dag}) |\phi|^2,
\eeq
where $A$ is the string theoretic axion and $\phi$ is a linearized QCD axion multiplet.
Assuming $\partial_{A}\log(Z) ={\cal O}(1)$, the kinetic mixing and the effective
decay constant are roughly given by $\epsilon \sim f_a/f_H$ and $f_{\rm eff} \sim f_H$.
The hidden axion mass $m_H$ can be generated through hidden non-perturbative effects 
such as  brane instantons. Therefore such a scenario is similar to the case discussed 
in Ref.~\cite{Ishida:2014dlp}, except for the presence of the (field theoretic) QCD axion.
If the saxion (${\rm Re}[A]$) dominates the Universe, it will decay into the SM particles
via the kinetic mixing with the QCD saxion. If the inflation scale is so large as suggested by BICEP2, 
the isocurvature constraint on the string axion is so tight that it is difficult to reconcile the
tension with the $3.5$\,keV X-ray line~\cite{Kawasaki:2014una}, unless $f_{\rm eff}$ and therefore
the string scale is lowered to allow $r_d \ll 1$~\cite{Kawasaki:2014una}.\footnote{Late-time entropy 
production by thermal inflation may solve both the tension and the cosmological moduli problem 
of the light moduli field in LVS.\cite{Choi:2012ye}}

\section*{Acknowledgment}
This work was supported by the Grant-in-Aid for Scientific Research on
Innovative Areas (No.23104008 [FT]),  JSPS Grant-in-Aid for
Young Scientists (B) (No.24740135 [FT] and No. 25800169 [TH]), Scientific Research (B) (No.26287039 [FT]), 
Inoue Foundation for Science [FT].  This work was also
supported by World Premier International Center Initiative (WPI Program), MEXT, Japan [FT].

\end{document}